\documentstyle[aps,eqsecnum]{revtex} 
\draft
\twocolumn
\begin{document}

\title{Atomic position localization via dual measurement}
\author{Hyunchul Nha$^{1,2}$, Jai-Hyung Lee$^{1}$, Joon-Sung Chang$^{1}$, and Kyungwon An$^{2}$}
\address{$^1$Department of Physics, Seoul National University, Seoul, Korea} 
\address{$^2$Center for Macroscopic Quantum-Field Lasers and Department of Physics, KAIST, Taejon, Korea} 
\date{\today}
\maketitle

\begin{abstract} 
We study localization of atomic position when a three-level atom interacts with a quantized standing-wave field in the Ramsey interferometer setup. Both the field quadrature amplitude and the atomic internal state are measured to obtain the atomic position information.  It is found that this dual measurement scheme produces an interference pattern superimposed on a diffraction-like pattern in the atomic position distribution, where the former pattern originates from the state-selective measurement and the latter from the field measurement. 
The present scheme results in a better resolution in the position localization than the field-alone measurement schemes.  We also discuss the  measurement-correlated mechanical action of the standing-wave field on the atom in the light of Popper's test.
\end{abstract}

\section{Introduction}
Atomic position localization has been an intriguing subject from the early days of the quantum mechanics, as seen in Heisenberg's microscope,  a well-known thought experiment devised for illustrating the uncertainty principle\cite{Heisen}.  Heisenberg's microscope exploits the interaction of an atom with light: The information on atomic position is obtained by detecting the scattered light. The resolution is limited roughly to a half wavelength of the light due to the wave nature of light. The uncertainty in atomic momentum automatically arises since the light imparts a mechanical momentum to the atom \cite{Braginsky}.

In a modern version of Heisenberg's microscope, on the other hand, one considers a quantum or classical light field with "standing-wave" mode structure in order to localize the atom without scattering photons \cite{storey,Kunze,Fam,sajid}. Because the strength of the interaction in a standing-wave field depends on the position, the observable quantities such as the phase shift of the atomic dipole, or that of the light field vary according to the atomic position. Thus, the measurements of these quantities yield information on the atomic position. All these schemes can determine the atomic position only within one period (i.e. a half wavelength) of the standing-wave structure due to the translational symmetry of the interaction.

Recently Storey {\it et al.} have proposed to measure the quadrature amplitudes of light interacting with a two-level atom at far-off resonance\cite{storey}. They also suggested that changing the width of a 'virtual slit', produced in the measurement of the field, by varying the phase of the field being measured, can implement Popper's test\cite{popper}. On the other hand, the localization by the measurement of the phase shift of the atomic dipole moment in a Ramsey interferometer set up was proposed using the classical standing-wave field\cite{Kunze}, or the quantum field\cite{Fam}, and was demonstrated experimentally by Kunze {\it et al.}\cite{kunze1}. Atomic position can also be localized by the frequency measurement of the photons scattered spontaneously from the atom\cite{sajid}, using the fact that the Mollow-sideband spectrum in the resonance fluorescence depends on the atomic position in the standing wave field. Other techniques, which do not involve the interaction with the standing-wave field, were proposed such as atom imaging methods\cite{thomas}, where an inhomogeneous magnetic field\cite{Stokes} or light intensity\cite{Gardner} causes a spatially-varying atomic level-shift, which correlates the atomic resonance frequency with the atomic position.

Until now, only a single observable among the various physical quantities involved in the atom-field dynamics has been chosen to be measured in order to localize the atomic position. In this work, we consider localization by {\it dual measurements}, i.e., we measure two observables for the localization of the atomic position. Our motivation for this scheme is twofold. First, it is based on the naive expectation that the more observables we measure, the better localization we could get. Of course, these variables should be correlated with the atomic position. Otherwise, increasing the number of the measured observables would be of no use. 

Our second motivation is more academic in that we hope to investigate the dynamics of the system
conditioned {\it a posterior} on the measurements. 
 Let us denote the Hermitian operators of the measured observables by $\hat A$,$\hat B$ ([$\hat A$,$\hat B$]=0) and their eigen-values by $a_i,b_i$ $(i=1,2,\cdots)$, respectively. We represent the system state before measurement by the wavefunction $|\Psi\rangle=\Sigma_ic_i|a_i\rangle\otimes|\Psi_i\rangle$ and assume that the value $a_m$ is obtained by the $\hat A$ measurement. (The states $|\Psi_i\rangle$ belong to the Hilbert spaces independent of the operator $\hat A$.) Then, the state after the measurement, $|\Psi_m\rangle$, is obtained by performing the projection operator $|a_m\rangle\langle a_m|$ to $|\Psi\rangle$. The state $|\Psi_m\rangle$ can be decomposed in terms of the bases
$|b_i\rangle$, i.e., $|\Psi_m\rangle=\Sigma_jd_j|b_j\rangle\otimes|\phi_j\rangle$. If one measures further the observable $\hat B$, then each wavefunction $|\phi_i\rangle$ (independent of $\hat A$ and $\hat B$) can be retrieved. However, if the observable $\hat B$ is not measured, the data collected through only $\hat A$ measurement would yield the state of the system traced over the observable $\hat B$, only to lose information of each wavefunction $|\phi_j\rangle$. The dynamics conditioned {\it a posterior} on the measurement ($\hat B$ measurement in this case) could demonstrate a substantially distinguished phenomenon compared with the usual averaged one\cite{car1}.

In this paper, we localize the atomic position by measuring both the quadrature amplitude of the light and the internal state of the atom.
We consider a three-level atom interacting with a single quantized standing-wave field in a Ramsey interferometer setup.
The atom has two hyperfine ground levels $|a\rangle$,$|b\rangle$ and one excited level $|c\rangle$. The quantized field couples both the states $|a\rangle$ and $|b\rangle$ to $|c\rangle$ at far-off resonance with different coupling constants. The interaction Hamiltonian shows that both the field state and the internal states are entangled with the atomic position. 

In addition to the position localization, we investigate the mechanical action of the standing-wave field on the atom, introduced inevitably by the localization. We show that the momentum uncertainty caused by the mechanical action does not hinder the Popper's test in the localization schemes such as the one proposed by Storey {\it et al.}\cite{storey} and that in a certain situation, the uncertainty can be minimized. 

This paper is organized as follows. 
In Section II, we introduce the interaction Hamiltonian and calculate the final state obtained after the interaction in the Raman-Nath regime. We show in Section III that the localization by the measurements, in the near region, i.e. right after the interaction, yields a position distribution which resembles an interference pattern superimposed on a diffraction pattern.
The interference pattern is attributed to the measurement of the atomic internal state while the diffraction to that of the quadrature amplitude of the light.
It is found that the localization is sharper than in the single measurement schemes, so the uncertainty $\Delta x$ of the position localization is smaller. 
In Section IV, the position distribution in the far region is presented and it is explained in terms of the dipole force by the standing-wave field correlated with the measurements. Especially, we are interested whether Popper's test can be implemented in the  measurement schemes exploiting the interaction with the standing-wave field in Section V. 
We summarize the results in section VI.

\section{System setup and the Hamiltonian}
In Fig.\,1, the system configuration is depicted with the appropriate atomic level diagram. The atom is initially prepared in the state $|a\rangle$. The atom first enters the microwave-field region ($\pi/2$-pulse), where the internal state is transformed to the superposed one $(|a\rangle+|b\rangle)/\sqrt{2}$. Next, it interacts with the standing-wave field inside the cavity at far-off resonance. The cavity frequency $\omega_C$ is tuned in the midway between
the atomic transition frequencies $\omega_{ca}$, $\omega_{cb}$ so that $\delta_a=\omega_C-\omega_{ca}<0$ and $\delta_b=\omega_C-\omega_{cb}>0$. If both the detunings satisfies the conditions $|\delta_a|\gg\Gamma_a$, $|\delta_b|\gg\Gamma_b$, where $\Gamma_a,\Gamma_b$ are the spontaneous emission rates from the excited state $|c\rangle$, then the excited state is rarely populated through the interaction, and thus can be ignored. For example, the Rb$^{85}$ atom has two hyperfine ground levels F=2,3 with the splitting $2\pi\times3$ GHz and the decay rates roughly $2\pi\times6$ MHz, so the conditions $|\delta_a|\gg\Gamma_a$ and $|\delta_b|\gg\Gamma_b$ are readily satisfied.

By the method of the adiabatic elimination\cite{Cohen-Tannoudji}, the interaction Hamiltonian inside the cavity is then written in the form as
\begin{eqnarray}
H=\hbar\sin^2(k_0\hat x)(g_a^2/\delta_a|a\rangle\langle a|+g_b^2/\delta_b|b\rangle\langle b|)a^{\dag}a
\end{eqnarray}
where $g_a$ and $g_b$ are the vacuum Rabi-frequencies associated with the transitions $|a\rangle\leftrightarrow|c\rangle$ and $|b\rangle\leftrightarrow|c\rangle$, respectively, $a^{\dag}$($a$) is the photon creation (annihilation) operator for the cavity mode and $k_0=2\pi/\lambda$ is the wave vector of the standing-wave mode. We set $-g_a^2/\delta_a=g_b^2/\delta_b\equiv G>0$, which can be adjusted by controlling the cavity frequency appropriately. We neglect the kinetic energy term in the Raman-Nath approximation, where the interaction time is short enough for the motional effect to be neglected. 

Let us denote the initial position distribution by $f(x)$ and assume that the cavity mode is in a coherent state $|\alpha\rangle$.
If the atomic internal state is $C_a|a\rangle+C_b|b\rangle$ before entering the cavity ($C_a=C_b=1/\sqrt{2}$ in our case), the initial state of the total system is denoted by
\begin{eqnarray}
|\Psi_0\rangle=\int dx f(x)|x\rangle\otimes (C_a|a\rangle+C_b|b\rangle)\otimes|\alpha\rangle
\end{eqnarray}
After interacting with the standing wave cavity-field for time $\tau$, the state becomes
\begin{eqnarray}
|\Psi(\tau)\rangle &=& e^{-iH\tau/\hbar}|\Psi_0\rangle=e^{iG(\hat x)\tau a^{\dag}a
(|a\rangle\langle a|-|b\rangle\langle b|)}|\Psi_0\rangle\nonumber\\
&=&\int dx f(x)|x\rangle\otimes[C_a|a\rangle\otimes|\alpha e^{iG(x)\tau}\rangle \nonumber\\
& &+C_b|b\rangle\otimes|\alpha e^{-iG(x)\tau}\rangle],
\end{eqnarray}
where 
\begin{eqnarray}
G(x)=G\sin^2(k_0 x).
\end{eqnarray}
In Eq.\,(3), we see that the phase of the cavity field is altered through the interaction. If the atomic internal state is $|a\rangle$, then the field state is rotated in the Wigner diagram (see Fig.\,2) by the angle 
\begin{eqnarray}
\Theta_a(x)=G(x)\tau.
\end{eqnarray} 
On the other hand, if the atom is in state $|b\rangle$ , the rotation angle is given by
\begin{eqnarray}
\Theta_b(x)=-G(x)\tau.
\end{eqnarray} 
Note that the field state after the interaction is entangled with the internal state as well as the atomic position, as seen in the above Eqs.\,(5) and (6). 
If the quadrature amplitude $X_{\theta}=ae^{-i\theta}+a^{\dag}e^{i\theta}$ of the field is measured, the atomic position is then localized.
We set $G\tau=\pi$ and select $\theta=0$ so that the measured quantity $X_0$ corresponds to the x-axis value in the Wigner diagram.
When the atom is localized at the node ($k_0x=0$), between the node and the antinode ($k_0x=\pi/4$), and at the antinode ($k_0x=\pi/2$),
the field state is accordingly rotated by the angles $0,\pm\pi/2,\pm\pi$, respectively, where $\pm$ sign refers to the case of the internal state $|a\rangle$ and $|b\rangle$, respectively. Then, the measured amplitude $X_0$ will be roughly $2\alpha,0,-2\alpha$.
Conversely, if the field amplitude is measured to be $2\alpha,0,-2\alpha$, then the atom is localized at the node, between the node and the antinode, and at the anti-node, respectively, via the measurement. This is how the localization by the field measurement comes about\cite{storey}.

After interacting with the cavity field, the atom enters the second $\pi/2$-pulse microwave region. Thus, the final state is given by
\begin{eqnarray}
|\Psi_{final}\rangle&=&\int dx f(x)|x\rangle\otimes[1/2|a\rangle\nonumber\\
& &\otimes(|\alpha e^{iG(x)\tau}\rangle-|\alpha e^{-iG(x)\tau}\rangle)\nonumber\\
& &+1/2|b\rangle\otimes(|\alpha e^{iG(x)\tau}\rangle+|\alpha e^{-iG(x)\tau}\rangle)].
\end{eqnarray}  
In the following, we use the formula
\begin{eqnarray}
|\chi_\theta\rangle=\frac{1}{^4\sqrt{2\pi}}\exp[-\frac{1}{2}(a^{\dag}e^{i\theta}
-\chi_\theta)^2+\frac{1}{4}\chi_{\theta}^2]|0\rangle \;,
\end{eqnarray}
where $|\chi_\theta\rangle$ is the eigenstate of the operator $X_{\theta}$\cite{storey} with $X_{\theta}|\chi_{\theta}\rangle=\chi_{\theta}|\chi_{\theta}\rangle$, and thus
the inner product $\langle\chi_{\theta}|\alpha e^{i\eta}\rangle$ is given by
\begin{eqnarray}
\langle\chi_{\theta}|\alpha e^{i\eta}\rangle=e^{-\left[(\alpha_r-\frac{\chi_{\theta}}{2})^2+i\alpha_i(\alpha_r-\chi_{\theta})\right]}\;,
\end{eqnarray}
where $\alpha_r={\rm Re}[\alpha e^{i(\eta-\theta)}]$ and $\alpha_i={\rm Im}[\alpha e^{i(\eta-\theta)}]$.

Now, if the field state is measured to be found in the eigen-state $|\chi_0\rangle$ after the entire interaction, it is found from Eqs.\,(7) and (9) that the the system collapses to the state 
\begin{eqnarray}
|\Psi_c\rangle\propto\int dx f(x)|x\rangle\otimes[D(x)I_a(x)|a\rangle+D(x)I_b(x)|b\rangle]\;,
\end{eqnarray}
where
\begin{eqnarray}
D(x)=\exp[-(\alpha\cos(G(x)\tau)-\chi_0/2)^2]
\end{eqnarray}
and
\begin{eqnarray}
I_a(x)=-i\sin\Delta(x),\hspace{1cm}I_b(x)=\cos\Delta(x)
\end{eqnarray}
with
\begin{eqnarray}
\Delta(x)=\alpha\sin(G(x)\tau)(\alpha\cos(G(x)\tau)-\chi_{0})\;.
\end{eqnarray}
If we measure the atomic internal state in addition, the final position distribution of the atom is given by
\begin{eqnarray}
P_a(x)\propto|f(x)|^2 |D(x)I_a(x)|^2
\end{eqnarray}
when the atom is found to be in $|a\rangle$ and
\begin{eqnarray}
P_b(x)\propto|f(x)|^2 |D(x)I_b(x)|^2
\end{eqnarray}
in $|b\rangle$.
Thus, $F_a(x)=|D(x)I_a(x)|^2$ and $F_b(x)=|D(x)I_b(x)|^2$ may be interpreted as the filter functions for the initial position distribution
associated with our measurement scheme.

\section{Localization in the near region}
To understand where the filter functions $D(x),I_a(x),$ and $I_b(x)$ originate from, we now consider the case that the Ramsey fields are turned off
and the atom remains in the same internal state throughout the entire interaction. We then measure the quadrature amplitude of the field only, as in the scheme proposed by Storey {\it et al.}\cite{storey}.
If the atomic state is $|a\rangle$ and the field state is measured to be $|\chi_0\rangle$, then the final atomic state is given by
\begin{eqnarray}
|\Psi_c\rangle^a\propto\int dx f(x)|x\rangle\otimes D(x)e^{-i\Delta(x)}|a\rangle \;.
\end{eqnarray}
Similarly, if the atom is in $|b\rangle$, the state is 
\begin{eqnarray}
|\Psi_c\rangle^b\propto\int dx f(x)|x\rangle\otimes D(x)e^{i\Delta(x)}|b\rangle \;.
\end{eqnarray}
Note that the exponential $e^{\mp i\Delta(x)}$ in the integrand has different argument according to the internal state.
In this case, the final position distribution is the same regardless of the internal state, given by
\begin{eqnarray}
\Pi_a(x)=\Pi_b(x)\propto|f(x)|^2 D^2(x)
\end{eqnarray}
We see that the field measurement alone produces the amplitude filter $D(x)$ with the phase filter $\pm\Delta(x)$.

In our case, on the other hand, with the Ramsey fields turned on, the quantum interference occurs. If the final state is found to be in $|a\rangle$, the possible quantum paths are
$|a\rangle\rightarrow|a\rangle_c\rightarrow|a\rangle$ and $|a\rangle\rightarrow|b\rangle_c\rightarrow|a\rangle$, where  $|a,b\rangle_c$ denotes the atomic state inside the cavity. ( Recall that we assumed the initial state to be $|a\rangle$ in deriving Eq.\,(10).)
Because these two paths are indistinguishable, they interfere to give $e^{-i\Delta(x)}-e^{i\Delta(x)}\propto \sin\Delta(x)$.
Similarly, if the atom is found finally in $|b\rangle$, two paths $|a\rangle\rightarrow|a\rangle_c\rightarrow|b\rangle$ and $|a\rangle\rightarrow|b\rangle_c\rightarrow|b\rangle$ interfere to give $e^{-i\Delta(x)}+e^{i\Delta(x)}\propto \cos\Delta(x)$.
The different signs $\pm$ of the interference in the two cases can be traced back to Eqs.\,(3) and (7).
Thus, the additional filters $I_a(x),I_b(x)$ in Eq.\,(12) are produced by the quantum interference of the indistinguishable paths.

In Fig.\,(3), we plot $F_a(x)=|D(x)I_a(x)|^2$ and $F_b(x)=|D(x)I_b(x)|^2$ for the cases $\chi_0=\pm2\alpha,0$. The envelopes in the figures are given by $|D(x)|^2$, which would be produced by the field measurement alone without the Ramsey-fields. We see that the overall shapes  resemble an interference pattern ($|I_a(x)|^2, |I_b(x)|^2$) superimposed on a diffraction pattern ( $|D(x)|^2$). Such shape is analogous to the distribution in the far-field region of a two-slit 
interferometer with each slit having finite width. The variances $\Delta x$ of $F_a(x)$ and $F_b(x)$ are smaller than that of $|D(x)|^2$. Thus, the localization is improved by the dual measurement.

\section{Localization in the far region}
The localization in the near-region by the dual measurement may be somewhat difficult to compare with that by the field-measurement of alone, due to the limited resolving power of usual atom detectors (e.g., hot-wire detector).  Note that the localizations in the two different schemes are  discernible only in the sub-wavelength scale. However,
the results can be clearly distinguishable in the far-region position distribution. We consider first the case in which the atom initially has the flat-top distribution, i.e., 
\begin{eqnarray}
f(x)&=&{\rm const.} \hspace{2cm}|x/\lambda|\le1\nonumber\\
&=&0 \hspace{2.5cm}{\rm otherwise.}
\end{eqnarray}
In Fig.\,4, we plot the momentum distribution right after the localization by the dual measurement in the near region, which corresponds to the position distribution in the far-region. We see that the distribution $P_a(p)$ with the atom in state $|a\rangle$ is complementary to $P_b(p)$ with the atom in $|b\rangle$ state, that is, they are out of phase with respect to each other. The spacings in the distributions $P_a(p)$ and $P_b(p)$ are the same, $4\hbar k$, because both the filter functions $F_a(x)$ and $F_b(x)$ have the $\lambda/4$-periodic structure (See Fig.\,3). Note that the sum of $P_a(p)$ and $P_b(p)$ gives exactly $\Pi(p)$, the distribution which results from the field-measurement alone without the Ramsey-fields. Also note that
$\Pi(p)$ does not depend on the internal state of the atom in this case.
At first sight, these results may seem quite natural, but they are not always the case, as seen below. 
\subsection{Dipole force induced by the field measurement}
The localization in the measurement scheme using the standing-wave field is always accompanied by the mechanical action on the atom. This is because the standing-wave field at far-off resonance
exerts the dipole force to the localized atom. When the cavity mode is in the coherent state $|\alpha\rangle$, the potential which the atom experiences is roughly given by 
\begin{eqnarray}
U_{a,b}(x)=\mp\hbar G\alpha^2\sin^2(k_0\hat x),
\end{eqnarray}
with $a^{\dag}a\rightarrow\alpha^2$ inserted into Eq.\,(1), where $-(+)$ sign refers to the case of the internal state $|a\rangle$ ($|b\rangle$). 

To see the effect of the dipole force more clearly, let us assume that the initial distribution $f(x)$ is a well-localized Gaussian. In Fig.\,5, for example, $f(x)$ is assumed to be 
\begin{eqnarray}
f(x)=\frac{1}{{(2\pi\sigma^2)}^{1/4}}\exp\left[-\frac{(x-x_0)^2}{4\sigma^2}\right]
\label{eq21}
\end{eqnarray}
with $x_0/\lambda=1/4$ ( located in the midway between the node and the anti-node) and $\Delta x=\sigma=0.1\lambda$. It becomes more localized by the measurement with $\chi_0=0$ as shown in the figure. The solid (dotted) line corresponds to the case that the atom is measured to be in $|a\rangle$ ($|b\rangle$) with the Ramsey-fields on. 
Now, when the far-region distribution is observed, $\Pi(p)$ in the single measurement scheme without the Ramsey-fields is not the same as the sum of $P_a(p)$ and $P_b(p)$ any more. Moreover, $\Pi(p)$ does depend on the internal state of the atom. We can explain such distributions in terms of the dipole force correlated with the position localization.

To begin with, we first consider the case of the field measurement alone without the Ramsey-fields [Fig.\,5 (c)]. The localized position distribution in this case has largely three parts (not shown), similar to Fig.\,5 (a). If the atom enters the cavity region in state $|a\rangle$, it experiences the potential given by $U_a(x)$. The two small outer parts in the distribution experience the force in the negative $x$ direction,
and thus are brought together to generate the interference pattern in the far-region due to the coherence of the initial Gaussian wavepacket. The large central part experiences the force in the opposite direction to make the diffraction pattern in the far-region. The peak position of the momentum distribution marked in the figure can be calculated using Eq.\,(22) in the next section as $p_t=\hbar k_0G\tau\alpha^2\approx19.6\hbar k_0$ with $G\tau=\pi,\alpha=2.5$ and $k_0x=\pi/4$.  Similar argument can be given to the 
atom in $|b\rangle$. This explains the distributions $\Pi_a(p)$ and $\Pi_b(p)$ [See Fig.\,5 (c)].

On the other hand, when the Ramsey-fields are turned on, the internal state inside the cavity is the superposition of $|a\rangle$ and $|b\rangle$. Thus, all the three parts in the position distribution of Fig.\,5 (a) experience the force both in the positive and negative directions simultaneously. This results in the interference patterns in the left and the right side of the far-region distributions [See Fig.\,5 (b)]. The distributions $P_a(p)$ and $P_b(p)$ are different from each other since they result from different superposition of indistinguishable quantum paths, as explained in Section III. Of course, the sum of $P_a(p)$ and $P_b(p)$ is the same as that of $\Pi_a(p)$ and $\Pi_b(p)$.

In this manner, the far-region distributions in all cases can be explained in terms of the dipole force correlated with the localization
by the measurement. 

\section{Popper's test}
In this section, we explore whether the present dual measurement scheme can implement Popper's test\cite{popper}. Popper's test is intended to answer the question whether the knowledge 
of the position itself can increase the momentum uncertainty {\it without mechanical momentum transfer} (Copenhagen interpretation), contrary to the Heisenberg microscope.
 
To this end, Storey {\it et al.} proposed that changing the width of the 'virtual slit', produced in the measurement of the field, by varying the phase of the field being measured, implement Popper's test\cite{storey}. In their scheme one compares the result of
the X-quadrature measurement with that of the Y-quadrature measurement. Suppose that the two measurements yield different uncertainties $(\Delta x)_X>(\Delta x)_Y$ of the position distribution. If the far-region position distribution, which corresponds to the momentum distribution in the near-region, shows that $(\Delta p)_X<(\Delta p)_Y$ in the absence of any mechanical action, the Copenhagen interpretation is then proved. 
Even in the presence of the mechanical action, the above inequality of momentum uncertainty still proves the Copenhagen interpretation as long as the mechanical action does not favor the momentum inequality in the same way.
As we have seen in the previous section,
the localization exploiting the interaction with the standing-wave field inevitably introduces the mechanical action, i.e., the dipole force. Therefore, it is necessary to examine the mechanical action in more detail. 

\subsection{The distribution of the momentum transferred to the atom by the dipole force}
If the atom is located at the position $x$ like a point particle, the impulse given to the atom for time duration $\tau$ due to the dipole force is given by
\begin{eqnarray}
p_t(x)&=&{\rm force}\times\tau=-\nabla U(x)\times\tau\nonumber\\
      &=&\pm\hbar k_0 G\tau\alpha^2\sin(2k_0x)
\end{eqnarray}
Note that the transferred momentum $p_t$ depends on the position $x$. When the atomic position has a distribution other than the delta-function, then $p_t$ is not single-valued but distributed over some range. 
If the atom is localized by the measurement to have the probability density $P(x)$, the uncertainty of the momentum transferred by the dipole force, $\langle(\Delta p_t)^2\rangle=\langle p_t^2\rangle-\langle p_t\rangle^2$, 
 is calculated as
\begin{eqnarray}
\frac{\langle(\Delta p_t)^2\rangle}{(\hbar k_0)^2}=(&G&\tau\alpha^2)^2\left[\int dx P(x)\sin^2(2k_0x)\right.\nonumber\\
&-&\left.\left(\int dx P(x)\sin(2k_0x)\right)^2\right].
\end{eqnarray} 
Let us assume that the measurement localizes the atomic wavepacket as a Gaussian, like in Eq.\,(21), with $P(x)=|f(x)|^2$. Then, it is easily obtained that
\begin{eqnarray}
\frac{\langle(\Delta p_t)^2\rangle}{(\hbar k_0)^2}=\frac{(G\tau\alpha^2)^2}{2}(1&-&e^{-8(k_0\sigma)^2}\cos(4k_0x_0)\nonumber\\&-&2 e^{-4(k_0\sigma)^2}\sin^2(2k_0x_0))
\end{eqnarray} 
When the atom is localized well ($k_0\sigma\ll1$), Eq.\,(24) becomes
\begin{eqnarray}
\frac{\sqrt{\langle(\Delta p_t)^2\rangle}}{\hbar k_0}\approx&2(G\tau\alpha^2)(k_0\sigma)|\cos(2k_0x_0)|
\end{eqnarray}
The momentum uncertainty is smaller for the narrower distribution, as seen in Eq.\,(25), where $\sqrt{\langle(\Delta p_t)^2\rangle}$ is proportional to $\sigma$. Due to this fact, although the mechanical action is inevitably accompanied in the localization, Popper's test can  nevertheless be  implemented in the measurement schemes using the standing-wave field, as explained below.
 
The momentum uncertainty $\Delta p$ results from two sources if we follow the Copenhagen interpretation. One is the knowledge of the position ($\Delta p_k$) itself, and the other is the mechanical action by the dipole force in the standing-wave field ($\Delta p_t$ in Eq.\,(25)), so that $\Delta p\sim\Delta p_k+\Delta p_t$. 
Let us assume that two different measurement schemes (e.g., two different quadrature measurements in Storey's scheme) yield the position localization as $\Delta x_1>\Delta x_2$. 
Since the uncertainty caused by the dipole force is smaller in the narrower distribution, we have ($\Delta p_{t1}>\Delta p_{t2}$).
If the far-region distribution is broader in the case of the narrower localization, i.e. $\Delta p_1<\Delta p_2$, we should conclude that the momentum uncertainty caused by the "knowledge" is much larger in the narrower distribution, i.e., $\Delta p_{k1}<\Delta p_{k2}$.
Moreover, the uncertainty by the mechanical action of the dipole force can be eliminated when the measurement localizes the atom in the midway between the node and the anti-node ($\Delta p_t\approx0$ with $x_0/\lambda=1/4$ in Eq.\,(25)). This is also the case even when the position distribution $P(x)$ is not a Gaussian as assumed in Eq.\,(25) as long as the atom is well localized around at $x_0/\lambda=1/4$. Therefore, we conclude that the proposal by Storey {\it et al.} can implement Popper's test even when the mechanical action  due to the dipole force of the standing wave field is included in the localization.

\subsection{When the atomic internal state is the superposed one inside the cavity}
In the above, we considered the case that the atom is definitely in one internal state. When the Ramsey fields are turned on, the internal state of the atom inside the cavity is the superposed one. Then, even when the atom is treated like a point particle (i.e., $\sigma=0$ in Eq. (\ref{eq21})), a mechanical momentum uncertainty comes about because the mechanical momentum delivered to the atom in state $|a\rangle$  is in the opposite direction to the momentum delivered to the atom in state $|b\rangle$. For this reason the mechanical momentum uncertainty in the dual measurement, $\Delta p_{d,t}$, becomes always larger than that in the field measurement only, $\Delta p_{f,t}$. 
We have shown in section III that the uncertainty $\Delta x_d$ of the position localization in the dual measurement scheme
is less than $\Delta x_f$ in the field measurement only. However, even if we get the result $\Delta p_d > \Delta p_f$, this does not prove the Copenhagen interpretation since $\Delta p_{d,t} > \Delta p_{f,t}$, i.e., there is additional momentum uncertainty caused by the indefiniteness of the atomic internal state in our dual measurement scheme. 
Therefore, the comparison of the result of the dual measurement with that of the field measurement alone does not constitute Popper's test.

\section{Summary}
In this work, we have investigated the atomic position localization by the dual measurement, i.e., both the field and the atomic internal state measurements, compared to the field measurement alone. We have also discussed the mechanical action of the light field, correlated with the measurement, on the atom.  
We began by showing that the localization is improved by the dual measurement scheme compared with the field-only measurement scheme. We then showed that although the localization exploiting the interaction with the standing wave-field at far-off resonance inevitably brings about the momentum uncertainty by the mechanical action of the dipole force origin, Popper's test can be nevertheless implemented in the schemes in which the atomic internal state is definite inside the standing-wave field. Moreover, we have found that it is better to localize the atom in the midway between the node and the antinode in order to minimize the momentum uncertainty.

This work is supported by Creative Research Initiatives of the Korean Ministry of Science and Technology.

\begin{figure}
\caption{ The schematic diagram of the measurement setup with the energy levels of the atomic internal states. Here $\mu {\rm w}$  represents the microwave fields to induce transitions between the states $|a\rangle$ and $|b\rangle$. Both the quadrature amplitude of the field and the internal state of the atom are measured to obtain the position information.}
\end{figure}

\begin{figure}
\caption{ The Wigner diagram for the coherent state rotated by the angle according to the internal state through the interaction given by Eq.\,(1). The $x$-axis denotes the amplitude of quadrature $X_0=a+a^{\dag}$ and the $y$-axis that of $X_{\pi/2}=-i(a-a^{\dag})$ out of phase with $X_0$. The rotation angles $\Theta_a(x)$ and $\Theta_b(x)$ depend on the position $x$.}
\end{figure}

\newpage
\hbox{}

\begin{figure}
\caption{ The filter functions $F_a(x)=|D(x)I_a(x)|^2$ and $F_b(x)=|D(x)I_b(x)|^2$ for the position distribution caused by the dual measurement for cases $\chi_0=\pm2\alpha,0$ with $\alpha=2.5$. The envelopes (dotted lines) correspond to the single measurement of the field without the Ramsey-fields. The measurement of $\chi_0=2\alpha,-2\alpha,0$ of the field quadrature amplitude $X_0$ roughly localizes the atom at the node, at the antinode, and in the midway between the node and the antinode, respectively.}
\end{figure}

\begin{figure}
\caption{The momentum distributions after the dual measurement, which corresponds to the position distribution in the far-region, when the field quadrature is measured to be at $\chi_0=0$ with $\alpha=2.5$. 
The distributions $P_a(p)$ and $P_b(p)$ denote the distributions that are obtained when the internal state is measured to be in $|a\rangle$ and $|b\rangle$, respectively. 
For comparison, the momentum distribution $\Pi(p)$ in the single measurement of the field quadrature amplitude without the Ramsey-fields is presented. The initial position distribution ($f(x)$ in Eq.\,(9)) was assumed to be flat over the range $|x/ \lambda|\le 1$.}

\end{figure}

\begin{figure}
\caption{ (a) the probability density $P(x)$ of the localization by the dual measurement with the potential $U_{a,b}(x)$ drawn together. The solid (dotted) line corresponds to the case that the atom is measured to be in $|a\rangle$ ($|b\rangle$), with the field measured to be at $\chi_0=0$. The initial position distribution $f(x)$ was assumed as a Gaussian as in Eq.\,(21), with $x_0/\lambda=1/4$ and $\sigma=0.1\lambda$. On the right side, the momentum distributions $P(p)$ with the Ramsey-fields is plotted in (b) and $\Pi(p)$ without the Ramsey-fields  plotted in (c), respectively, corresponding to the localizations shown in (a). In Fig.\,(b) and (c), the solid/dotted lines are for the internal state $|a\rangle$/$|b\rangle$.}
\end{figure}


\begin{references}

\bibitem{Heisen} W. Heisenberg, Z. Phys. {\bf 43}, 172 (1927).
\bibitem{Braginsky} V. B. Braginsky and F. Y. Kahlili, {\it Quantum Measurement}, Cambridge University Press (1992).
\bibitem{storey} P. Storey, M. J. Collet, and D. F. Walls, \prl {\bf 68}, 472 (1992); P. Storey, M. J. Collet, and D. F. Walls, \prl {\bf 49}, 405 (1993).
\bibitem{Kunze} S. Kunze, G. Rempe, and M. Wilkens, Europhys. Lett. {\bf 27}, 115 (1994).
\bibitem{Fam} Fam Le Kien, G. Rempe, W. P. Schleich, and M. S. Zubairy, \pra {\bf 56}, 2972 (1997).
\bibitem{sajid} Sajid Qamar, S.-Y. Zhu, and M. S. Zubairy, \pra {\bf 61}, 068306 (2000).
\bibitem{popper} K. R. Popper, {\it Quantum Theory and the Schism in Physics} (Hutchison, London, 1982), pp. 27-29. The purpose of the Popper's test is to verify whether the Copenhagen interpretation of the uncertainty relation is correct or not.
\bibitem{kunze1} S. Kunze, K. Dickemann, and G. Rempe, \prl {\bf 78}, 2038 (1997).
\bibitem{thomas} J. E. Thomas, Opt. Lett. {\bf 14}, 1186 (1989); J. E. Thomas, \pra {\bf 42}, 5652 (1990).
\bibitem{Stokes} K. D. Stokes, C. Schnurr, J. R. Gardner, M. Marable, G. R. Welch, and J. E. Thomas, \prl {\bf 67}, 1997 (1991).
\bibitem{Gardner} J. R. Gardner, M. Marable, G. R. Welch, and J. E. Thomas, \prl {\bf 70}, 3404 (1993).
\bibitem{car1} One example is H. J. Carmichael, H. M. Castro-Beltran, G. T. Foster, and L. A. Orozco, \prl {\bf 85}, 1855 (2000).
\bibitem{Cohen-Tannoudji}
C.\,Cohen-Tannoudji, J.\,Dupont-Roc, G.\,Grynberg, {\it Atom-photon interactions: basic processes and applications},
John Wiley \& Sons, Inc. New York (1992).
\end{references}
\end{document}